\documentclass[12pt,preprint]{aastex} 

\usepackage{natbib}
\usepackage{epstopdf}
\usepackage{graphicx}
\usepackage{subfigure}
\usepackage{rotating}
\usepackage{textcomp,graphicx,amsmath,pdflscape,amssymb}

\begin{document}


\def\Msun{M_\odot}
\def\Lsun{L_\odot}
\def\Rsun{R_\odot}

\slugcomment{Submitted to ApJ}

\shorttitle{Globular Cluster Novae}
\shortauthors{Doyle et al}

\title{A {\it Hubble Space Telescope} Survey for Novae in the Globular Clusters of M87\altaffilmark{1}}

\author{Trisha F. Doyle\altaffilmark{2,3}, Michael~M.~Shara\altaffilmark{2}, Alec M. Lessing \altaffilmark{4}, and David~Zurek\altaffilmark{2}}


\altaffiltext{1}{Based on observations with the NASA/ESA {\it Hubble Space
Telescope}, obtained at the Space Telescope Science Institute, which is
operated by AURA, Inc., under NASA contract NAS 5-26555.}
\altaffiltext{2}{Department of Astrophysics, American Museum of Natural
History, Central Park West and 79th street, New York, NY 10024-5192, USA}
\altaffiltext{3}{Astroparticle Physics Laboratory, Goddard Space Flight Center, 8800 Greenbelt Road, Greenbelt, MD 20771, USA}
\altaffiltext{4}{Collegiate School, 301 Freedom Place South, New York, NY 10069, USA}

\begin{abstract}

The giant elliptical galaxy M87 has been imaged over 30 consecutive days in 2001, 60 consecutive days in 2005-2006, and every 5 days over a 265 day span in 2016-2017 with the {\it Hubble Space Telescope}, leading to the detection of 137 classical novae throughout M87. We have identified 2134 globular clusters (GC) in M87 in these images, and carried out searches of the clusters for classical novae erupting in or near them. One GC CN was detected in the 2001 data, while zero novae were found during the 2005-2006 observations. Four candidate GC novae were (barely) detected in visible light during the 2016-2017 observations, but none of the four were seen in near-ultraviolet light, leading us to reject them. Combining these results with our detection of one M87 GC nova out of a total of 137 detected CN, we conclude that such novae may be overabundant relative to the field, but small number statistics dominate this (and all other) searches. A definitive determination of GC CN overabundance (or not) will require much larger samples which LSST should provide in the coming decade.
\end{abstract}


\section{Introduction}

Classical novae (CN) are powered by thermonuclear runaways on the surfaces of white dwarf stars, which have accreted hydrogen-rich envelopes from red dwarf or red giant companions \citep{war95}. The luminosities of CN near maximum range from M$_V$ = -5.5 to -10.5 \citep{ayd18,sch18}, so they are detectable in galaxies all the way out to the Fornax and Virgo clusters. Surveys with large ground-based telescopes have searched for differences between the CN populations of spiral and elliptical galaxies; located intracluster tramp novae; and determined the nova rates in different kinds of galaxies \citep{pri87, nei05, cur15}.

CN should exist in all stellar populations, including those found in globular clusters (GC). The remarkable discovery that X-ray binaries are {\it enhanced} 100-fold in GC relative to the field \citep{kat75} inspired theory and models \citep{fpr75,kro84} of 100X enhanced cataclysmic variable \citep{dis94} and CN rates in GC. The fraction of stars, f, in globular clusters is {$\sim$} $2\times10^{-3}$ that of the stars in their host giant elliptical galaxies \citep{cur15}. The overall Galactic CN rate is {$\sim$} 50/yr, with an uncertainty of about 50\% \citep{sha17}. Assuming the same value of f in the Milky Way as in giant elliptical galaxies, one might expect 1600 CN/160 yr in our Galaxy's GC if the predicted 100X enhancement occurs.

In fact, only one certain CN (nova T Sco 1860 in M80) \citep{auw62,sha95} and one possible CN (Nova Oph 1938 in M14) \citep{saw38, wsb90} have been detected in the 150 GC of the Galaxy in the past 160 years, for an apparent Galactic rate of 4-8 x $10^{-5}$ CN/GC/yr. This corresponds to a factor of 8 {\it rarity} of Galactic GC CN relative to field CN. This ``rarity" is at least partly due to incompleteness. Most Galactic GC have not been monitored at least weekly for CN by amateur or professional astronomers since 1860, and the appearances of a bright star inside a GC could easily have been ignored. This realization has prompted searches for GC CN in external galaxies with much larger numbers of GC than the Milky Way, both with groundbased telescopes, the {\it Hubble Space Telescope} (HST) \citep{fer96, fer03, mad07, sha16} and {\it XMM-Newton} and {\it Chandra} \citep{hen13}. 

Over an effective survey time of 1 year, \cite{ctp90} and \cite{tcs92} monitored the entire M31 GC population in H$\alpha$, detecting zero CN. 
Three certain and one more possible CN have since been found in M31 \citep{sha07,hen09,pea10,cao12,hen13}. \citet{hen13} suggested that the M31 nova rate might be as high as 0.05 CN/GC/yr. \citet{sha04} found one GC CN during a 30 day survey of the GC of M87 with HST. Two additional GC CN were reported in GC of M84 and M87 by \citet{cur15}, who estimated a GC nova rate of 4x$10^{-4}$ CN/GC/yr. 

Because only nine GC CN are known, small number statistics dominate these searches. In addition, all ground-based searches to date are subject to significant incompleteness due to weather and lunar phases. Our ongoing survey of M87 with HST for CN \citep{sha16} partially overcomes these problems because HST observations rule out gaps due to weather, and there are no variations in limiting magnitude due to variable seeing or lunar phase. Our 2005-2006 daily imaging over a 9 week span, and 2016-2017 imaging every 5 days over a 265 day span, were deep enough to be almost impervious to M87's background light, revealing CN to within 10" of the galaxy's nucleus. In addition, field CN were detected over a nearly 6 magnitude range of brightness, so that even the faintest and fastest of M87 field novae were detected. Our datasets are sensitive to GC CN (see section 4.1), and we report the results of our latest searches here. In section 2 we summarize our HST observations of M87, and in section 3 we describe our search methodology for CN associated with its globular clusters. In section 4 we present our results, and we compare them with the rates of previous GC CN detections. Our conclusions are briefly summarized in section 5.
 
\section{Observations}

Over the 30 day interval 28 May 2001 through 25 June 2001 the {\it Hubble Space Telescope} Wide Field and Planetary Camera 2 (HST/WFPC2) provided imaging of the giant elliptical galaxy M87 in the F606W ($V$ band) and F814W ($I$ band) filters, taken for HST Cycle-9 program 8592 (PI - J. Silk). Four 260-second exposures in the F814W filter were followed by one exposure of 400 seconds in the F606W filter during each one-orbit epoch of observations. The 30 epochs were spaced $1.0 \pm 0.1 $ days apart. The observations, data reductions, detections and characterizations of M87 variables (mostly novae) are given in \citet{bal04}. We refer to these data as the ``2001 dataset". 

Over the 72 day interval 24 December 2005 through 5 March 2006 we carried out {\it Hubble Space Telescope} Advanced Camera for Surveys (HST/ACS) imaging of the giant elliptical galaxy M87 in the F606W ($V$ band) and F814W ($I$ band) filters, taken for HST Cycle-14 program 10543 (PI - E. Baltz). The observations, data reductions, detections and characterizations of the M87 novae we detected are given in \citet{sha16} (hereafter Paper I). Four 360-second exposures in the F814W filter were followed by one exposure of 500 seconds in the F606W filter during each one-orbit epoch of observations. The first four epochs were spaced $5.00 \pm 0.02 $ days apart, followed by $1.00 \pm 0.11$ day spacing for the remaining 60 epochs. Photometric errors in F814W range from 0.01 - 0.04 mag for M87 novae near maximum at 22nd magnitude to 0.3 - 0.4 mag by the time they have faded by 2 - 3 mag from maximum. We refer to these data as the ``2005-2006 dataset". 

Over the 265 day interval 13 November 2016  through 31 July 2017 we carried out {\it Hubble Space Telescope} Wide Field Camera3 (HST/WFC3) imaging of M87 in the F606W ($V$ band) and F275W ($NUV$ band) filters, taken for HST Cycle-24 program 14618 (PI - M. Shara). The observations, data reductions, detections and characterizations of the M87 novae we detected will be published elsewhere. Two 360-second exposures in the F606W filter were followed by three exposure of 500 seconds each in the F275W filter during each one-orbit epoch of observations. The 53 epochs were spaced $5.00 \pm 0.3 $ days apart. Photometric errors in F606W range from 0.01 - 0.04 mag for M87 novae near maximum at 22nd magnitude to 0.3 - 0.4 mag at 25th magnitude. We refer to these data as the ``2016-2017 dataset".

\section{Data Analysis}
\subsection{Globular Clusters}

We used the 2005-2006 dataset to create two median images, one of the F814W images, and one of the F606W data. DAOFIND \citep{ste87} within pyRAF was applied to each image to generate a list of candidate globular clusters. The sky image output of DAOFIND was blurred to create a background galaxy image, which we subtracted from the median images. These sky subtracted images were used to eliminate candidate GC with significantly elongated or irregular radial profiles. Our refined list of 2134 globular clusters display V magnitudes between 21 and 26 and V-I colors between 0.5 and 1.5 (we conservatively excluded sources outside these limits). The overlap with the 2250 objects in the M87 GC list of \citet{pen09} is excellent, with their extra GCs almost all displaying V magnitudes fainter than 26.0.

\subsection{Nova Searches}

We used the package ASTRODRIZZLE to combine the four F814W images at each epoch of the 2005-2006 dataset. We used PHOT from pyRAF to obtain photometry of every GC at each epoch. For each GC we identified epochs of interest as those in which a globular cluster brightened by at least 3 sigma relative to the mean brightness it exhibited in the other epochs. Each such candidate's 4 F814W images were examined individually to eliminate obvious cosmic rays or other sources of noise. No nova candidates were found in the 2005-2006 dataset.

We used ASTRODRIZZLE to combine the 2016-2017 dataset images for each epoch in each of the F275W and F606W images. Because cosmic rays are common and we had only two F606W images per epoch, we used the software's minmed parameter to create a median image, conservatively setting the sigma parameters such that the software would always choose the minimum image. Next, we used PHOT to obtain photometry for these minimum F606W images at each epoch. For each GC we identified epochs of interest as those in which a globular cluster brightened by at least 3 sigma relative to the mean brightness it exhibited in the other 52 epochs. Each such candidate's 5 images (two in F606W and three in F275W) were examined individually to eliminate obvious cosmic rays or other sources of noise. Four nova candidates, each a few tenths of a magnitude brighter than its host globular cluster, were detected in the minimum image F606W frames. Their brightnesses (after subtracting off their GC hosts), ranged from m(F606W) = 23.5 to 24.8 . Using the distance modulus to M87 derived by \citet{bir10}, i.e. (m-M)$_0 = 31.08 \pm 0.06 $, these candidate nova brightnesses correspond to absolute I-band magnitudes of -7.6 to -5.3, typical of M87 CN near maximum light \citep{sha16}. None of the four candidates showed a significant detection in any of the F275W frames for the entire duration of the 9 months of the 2016-2017 observations. 

CN usually peak in ultraviolet brightness a few days before \citep{cao12}, or days to weeks after visual maximum (e.g. \citep{gal74,cao12}, with UV minus Visual colors close to zero. Three of the four candidates were bright enough to have been detected in our F275W frames. That none were detected leads us to discard them as likely CCD or low level cosmic ray artifacts. A fourth candidate, seen in the minimum brightness image of the F606W image of 18 March 2017 at RA(2000) = 12:30:47.519, Dec (2000) = +12:23:18.11 was too faint to have been detected in F275W. Its uncontaminated detection in just one F606W CCD frame is insufficient for us to consider it to be real. In summary, we detected zero novae in 2134 GC of M87 in either of our (72 days-long) 2005-2006 or (265 days-long) 2016-2017 databases, in contrast with 1 GC CN detected during our earlier 30 day survey \citep{sha04}. 

\section{Results}
\subsection{Completeness}

Could we be missing most M87 GC novae because they are masked by the brightness of their host GC? To test our completeness we determined what fraction of synthetic novae added to the 2004-2005 F814W images and the 2016-2017 F606W images we could recover. We used IRAF's PSF function to generate a PSF model from the brighter stars in the field of the galaxy in each filter. For each GC, we randomly selected one or more epochs in which to add novae. For these epochs, we randomly selected a nova magnitude between 21 and 27 and a distance from each GC center between 0 and 5 pixels. We then used ADDSTAR to add simulated novae to the original images. Running the same nova candidate detection software on these images (with ${\sim}$5000 simulated novae in each filter), we created completeness curves for recovery of these simulated novae, shown in figures 1a and 1b. Both figures demonstrate that we can recover $\sim50\%$ of erupting GC CN that are within ${\sim}$ 2 magnitudes of their cluster hosts.

Half of all M87 GC are fainter than m(F814W) = 22.5  and m(F606W) = 23.5 \citep{pen09}, so we can detect CN in these GC that become brighter than m(F814W) ${\sim}$ 24.5 and  m(F606W) ${\sim}$ 25.5. 30 of 31 CN identified by \citet{sha16} in M87 became brighter than m(F606W) = 25.5, while 25 of the same 31 CN became brighter than m(814W) = 24.5. $\sim 20\%$ of M87 CN become as bright as m(606W) = 23, making them detectable in even the most luminous M87 GC. This demonstrates that our searches have detected at least $50\%$ of all GC CN, but significantly less than $100\%$.

\subsection{The M87 GC CN Rate}

The moderately different areal coverages of the three HST instruments used for these surveys (the Wide Field Camera 2 (WFPC2), the ACS and WFC3) mean that the number of GC checked for CN is slightly smaller in the WFPC2 survey. Ignoring this modest difference, our detection rate is  ${\sim}$ 1 CN/2000 GC/367 days ${\sim}$ 5x$10^{-4}$ CN/GC/yr, in good agreement with the rate measured by \citet{cur15}, 4x$10^{-4}$ CN/GC/yr. Those authors argued, on the basis of their discovery of two GC novae (one in M84, one in M87 and zero in M49) out of a total of 83 CN  that novae are likely enhanced relative to the field by at least an order of magnitude. While our rates are in apparently good agreement, we believe it premature to claim an enhanced nova rate in GC.

We detected ${\sim}$100 CN in M87 in the 2016-2017 dataset, 31 in the 2005-2006 dataset \citep{sha00}, and at least six in the 2001 dataset, for a total of 137 CN, compared to just one GC CN. Our detections of 1/137 GC CN, and those of \citep{cur15} (2/83) are dominated by small number statistics. Our small number statistics, and those of all other surveys, suggest to us that the Large Synoptic Survey Telescope's planned imaging of the Virgo and Fornax clusters of galaxies every few nights for several years, in the coming decade, will be essential to finally provide 100 or more GC CN, and thousands of galaxy CN. Only then will a definitive determination of the rate of GC CN in different galaxies, and a direct measurement (if real) of the overabundance of GC CN relative to those in the field, be possible.

\clearpage

\begin{figure}
\centerline{\includegraphics[width=0.99\columnwidth]{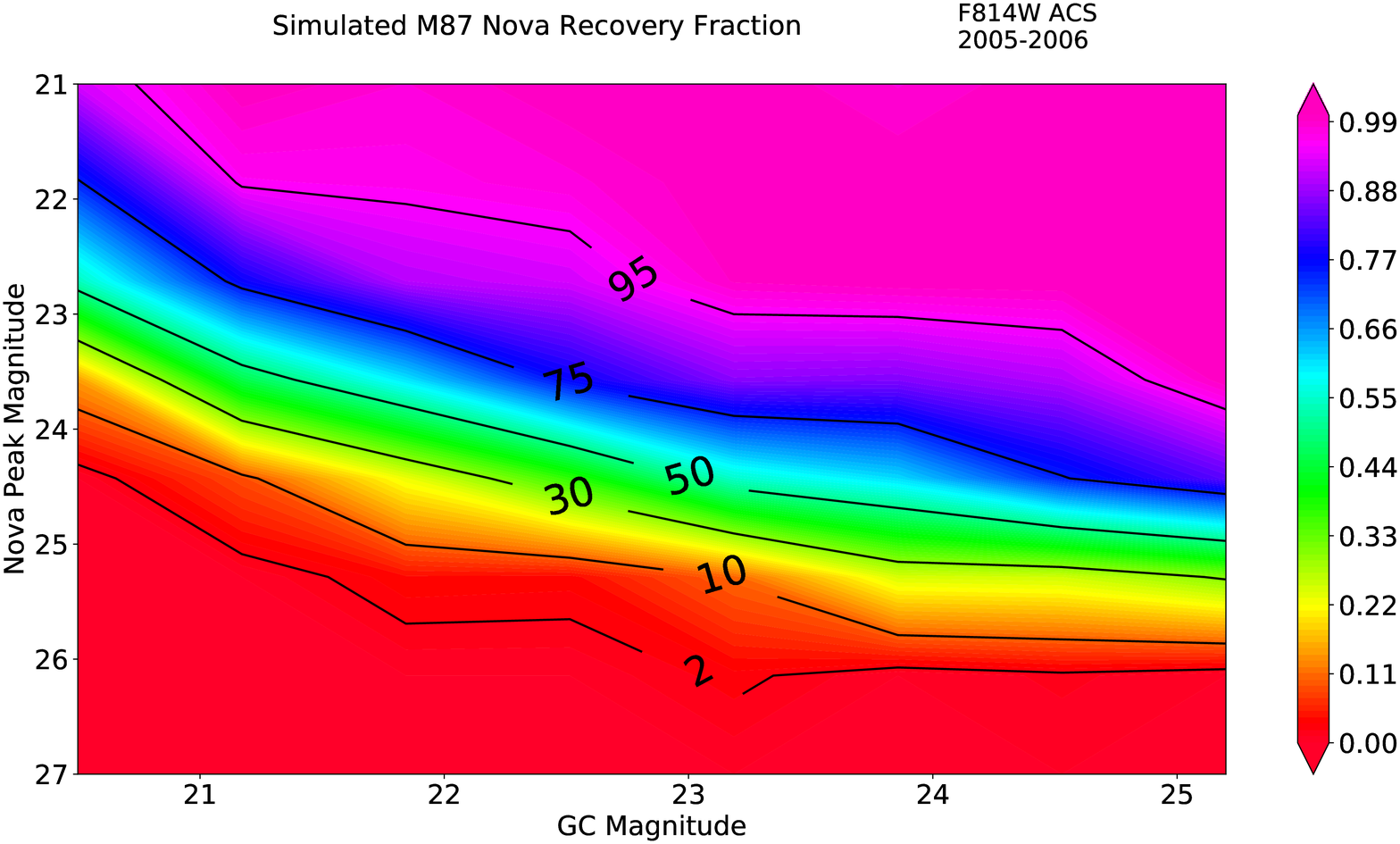}}
\centerline{\includegraphics[width=0.99\columnwidth]{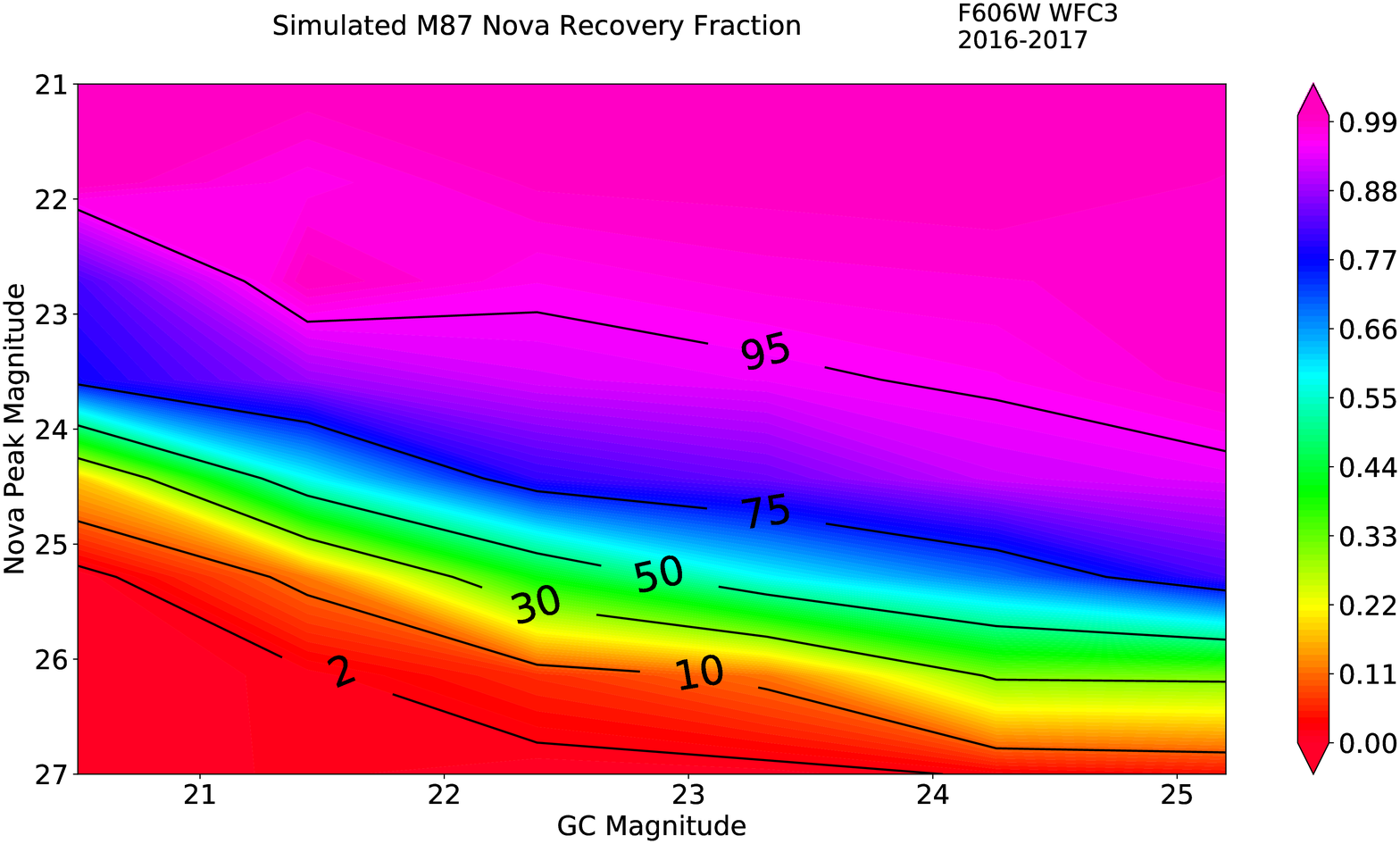}}
\caption{(Top) The completeness fraction of recovery of synthetic novae in globular clusters imaged with the HST ACS/F814W (I-band) filter.  
(Bottom) The completeness fraction of recovery of synthetic novae in globular clusters imaged with the HST WFC3/F606W (V-band) filter.}\label{spectra}
\end{figure}

\clearpage

\section{Summary}
We have searched HST images of ${\sim}$ 2000 GC of the Virgo giant elliptical galaxy M87, taken over a total timeframe of one year. One GC CN was observed to erupt during that time. Our search was sufficiently sensitive to have detected at least half of all erupting CN in GC. The rate we deduce is ${\sim}$ 5x$10^{-4}$ CN/GC/yr. Only much larger samples, expected in the coming decade from LSST, can definitively determine if this rate is correct, and if GCs produce CN at a higher rate than the field.

\acknowledgments

We gratefully acknowledge the support of the STScI team responsible for ensuring timely and accurate implementation of our M87 programs. Support for programs \#10543 and \#14618 was provided by NASA through a grant from the Space Telescope Science Institute, which is operated by the Association of Universities for Research in Astronomy, Inc., under NASA contract NAS 5-26555. PyRAF is a product of the Space Telescope Science Institute, which is operated by AURA for NASA.

\end{document}